\documentclass[11pt]{article}
\usepackage[a4paper,margin=25mm]{geometry}
\usepackage{amsmath,amssymb,bm}
\usepackage{xcolor}
\usepackage[T1]{fontenc}
\usepackage{lmodern}
\usepackage{microtype}
\usepackage[hidelinks]{hyperref}
\usepackage{cite}
\allowdisplaybreaks

\def\##1{{\bf #1}}
\def\=#1{\underline{\underline #1}}
\def\^#1{{\rR}eve{#1}}
\def\`#1{{#1^\prime}}
\def\:#1{#1^{\prime\prime}}

\def\.{\mbox{ \tiny{$^\bullet$} }}

\def\ler{\left(}
\def\rir{\right)}
\def\les{\left[}
\def\ris{\right]}

\def\xit{\xi_{\rm t}}
\def\zetat{\zeta_{\rm t}}
\def\epst{\eps_{\rm t}}
\def\mut{\mu_{\rm t}}
\def\epsz{\eps_{\rm z}}
\def\muz{\mu_{\rm z}}

\def\calH{{\cal H}}
\def\Heps{{\cal H}_\eps}
\def\Hmu{{\cal H}_\mu}
\def\Leps{{\cal L}_\eps}
\def\Lmu{{\cal L}_\mu}

\newcommand{\G}{\=G}
\newcommand{\E}{\=E}
\newcommand{\F}{\=F}
\newcommand{\La}{\=L_{\rm adj}}
\newcommand{\LL}{\={ L}_{ee}}
\newcommand{\M}{\=M}

\newcommand{\uv}[1]{\bm{u}_{#1}}
\newcommand{\xx}{\bm{r}}
\newcommand{\xp}{\bm{r}'}
\newcommand{\Id}{\=I}
\newcommand{\It}{\=I_{\rm t}}
\newcommand{\eps}{\varepsilon}
\newcommand{\dt}{\delta_{\rm t}}
\newcommand{\kt}{k_{\rm t}}
\newcommand{\nt}{\nabla_{\rm t}}
\newcommand{\dA}{(A_+-A_-)}

\newcommand{\cz}{(\nabla\times\uv z)}
\newcommand{\curlI}{\nabla\times\Id}
\newcommand{\pzz}{\frac{\partial^2}{\partial z^2}}
\newcommand{\ce}{\bigl(\epsz\mut^2+\muz\xit^2-\muz\dt\bigr)}
\newcommand{\cm}{\bigl(\muz\epst^2+\epsz\xit^2-\epsz\dt\bigr)}

\title{ 
\Large Closed-form dyadic Green functions for  transversely uniaxial bianisotropic medium with equal magnetoelectric scalars }
\author{\begin{minipage}{0.95\textwidth}
\centering 
Akhlesh Lakhtakia$^{1,2,*}$ and Tom G. Mackay$^{3}$\\[6pt]
{\small
$^{1}$NanoMM -- Nanoengineered Metamaterials Group,\\
Department of Engineering Science and Mechanics,\\
The Pennsylvania State University, University Park, PA 16802, USA\\[3pt]
$^{2}$School of Mathematics, University of Edinburgh,\\
Edinburgh EH9 3FD, UK\\[3pt]
$^{3}$School of Mathematics and Maxwell Institute for Mathematical Sciences,\\
University of Edinburgh, Edinburgh EH9 3FD, UK\\[6pt]
$^{*}$Corresponding author: \href{mailto:akhlesh@psu.edu}{akhlesh@psu.edu}}
\end{minipage}}
\date{}

\begin{document}
\maketitle

\begin{abstract}
\noindent
A transversely uniaxial bianisotropic medium with equal magnetoelectric scalars   was conceived by Weiglhofer in 1994. Using the scalar factorization method, he concluded that the dyadic Green functions of this medium could not be found in closed form. That conclusion is incorrect. The essentials of the derivation are reproduced here, and closed-form expressions are given for all four dyadic Green functions of the unbounded medium, linking the electric and magnetic fields to electric and magnetic source current densities.
\end{abstract}

\noindent\textbf{Keywords:} dyadic Green function; uniaxial bianisotropic medium; scalar factorization; duality

\section{Introduction}
In 1994, Weiglhofer conceived the transversely uniaxial bianisotropic medium (TUBM) as a special case of the most general uniaxial bianisotropic medium \cite{W1994}. Such a medium is the homogenized description of a composite of electrically small uniaxial inclusions whose axes lie randomly in a plane normal to a fixed direction 
\cite{WLM1993}. Its constitutive relations $\bm D=\=\eps\.\bm E+\=\xi\.\bm H$ and $\bm B=\=\zeta\.\bm E+\=\mu\.\bm H$ contain the  dyadics
\begin{equation}
\left.\begin{aligned}
\=\eps&=\epst\Id+(\epsz-\epst)\uv z\uv z\,,\qquad &
\=\xi&=\xit\left(\Id-\uv z\uv z\right)\\
\=\zeta&=\zetat\left(\Id-\uv z\uv z\right)\,,\qquad &
\=\mu&=\mut\Id+(\muz-\mut)\uv z\uv z 
\end{aligned}\right\}\,,
\label{conrels}
\end{equation}
where $\=\eps$ is permittivity dyadic, $\=\mu$ is the permeability dyadic, $\=\xi$ and $\=\zeta$ are the magnetoelectric dyadics, and
$\=I$ is the identity dyadic. The scalars $\xit$ and $\zetat$ capture the magnetoelectric properties of the TUBM.
 Throughout, the permittivity and permeability dyadics are assumed to be nonsingular, i.e., $\epst\epsz\ne0$ and $\mut\muz\ne0$.

Weiglhofer \cite{W1994} employed the scalar factorization method he developed with Lindell \cite{LW1994,WL1994a,WL1994b}
 to determine
the electric dyadic Green function for the TUBM in closed form. This method requires the factorization of a  fourth-order scalar operator $\calH$. Weiglhofer found
that factorization is possible only if
\begin{equation}
\xit=\zetat\,.
\label{equal}
\end{equation}
Despite that factorization, he  could not determine the electric dyadic Green function $\G_{ee}(\xx,\xp)$ in closed form. He conclude
that ``the transversally uniaxial bianisotropic medium with $\xi_t=\zeta_t$ becomes the first
known case in which a closed-form solution for the dyadic Green function is not
forthcoming despite successful factorization of the fourth-order operator $\calH$.''

While exploring the capabilities of ChatGPT-6.1~Sol~Light to implement the scalar factorization method, we found that
Weiglhofer's conclusion regarding the TUBM with equal magnetoelectric scalars is incorrect.
In this short paper the essentials of the derivation are reproduced, and closed-form expressions are given for the dyadic Green functions linking (i) the electric field to the source electric current density, (ii) the electric field to the source magnetic current density, (iii) the magnetic field to the source magnetic current density and (iv) the magnetic field to the source electric current density.

Vectors are in bold and dyadics are double underlined. The unit vectors of the Cartesian coordinate system are $\uv x$, $\uv y$ and $\uv z$. 
The identity dyadic is denoted by $\Id$ and $\It=\Id-\uv z\uv z$. The fields are time harmonic, and a factor $\exp(-i\omega t)$ is suppressed.

\section{Field equations}

 With the source electric current density denoted by $\bm J$ and the source magnetic  current density denoted by $\bm K$, the two  Maxwell curl equations are:
\begin{equation}
\label{Maxwellcurl}
\left.\begin{aligned}
\ler \curlI-i\omega\=\zeta \, \rir \.\bm E(\xx)-i\omega\=\mu\. \bm H(\xx)&=-\bm K(\xx)\\
i\omega\=\eps\. \bm E(\xx)+ \ler \curlI+i\omega\=\xi \, \rir\. \bm H(\xx)&=\bm J(\xx)
\end{aligned}\right\}\,.
\end{equation}
Their solutions in the unbounded medium are written as \cite{T1993,OL2002,FL2018}:
\begin{equation}
\left.\begin{aligned}
\bm E(\xx)&=i\omega\iiint\G_{ee}(\xx,\xp)\. \bm J(\xp)\,d^3r'+\iiint\G_{em}(\xx,\xp)\. \bm K(\xp)\,d^3r'\\
\bm H(\xx)&=\iiint\G_{me}(\xx,\xp)\. \bm J(\xp)\,d^3r'+i\omega\iiint\G_{mm}(\xx,\xp)\. \bm K(\xp)\,d^3r'
\end{aligned}\right\}
\end{equation}
The electric dyadic Green function $\G_{ee}$ is the same as $\G$ of Ref.~\citenum{W1994}. It satisfies
the dyadic differential equation
\begin{equation}
\label{Gee-equation}
\LL\. \G_{ee}(\xx,\xp)=\delta(\xx-\xp)\Id\,,
\end{equation}
where $\delta(\xx-\xp)$ is the Dirac delta and  the dyadic differential operator
\begin{equation}
\LL\equiv \ler \curlI+i\omega\=\xi \, \rir \. \=\mu^{-1}\. \ler \curlI-i\omega\=\zeta\, \rir-\omega^2\=\eps\, .
\end{equation}

\section{Essentials of the scalar factorization method}
If there exists an adjoint operator $\La$ defined by
\begin{equation}
\label{def-La}
\La \. \LL =\calH\Id \,,
\end{equation}
then the solution of Eq.~\eqref{Gee-equation}
is
\begin{equation}
\G_{ee}=\La W(\xx,\xp)\,,
\end{equation}
where the scalar super-Green function $W(\xx,\xp)$ satisfies the differential equation
\begin{equation}
\label{eq9}
\calH W(\xx,\xp)=\delta(\xx-\xp)\,. 
\end{equation}

For the TUBM with equal magnetoelectric
scalars (cf. Eqs.~\eqref{conrels} and \eqref{equal}), the scalar fourth-order operator \cite{W1994}
\begin{equation}
\calH={\Heps}{\Hmu}-{\Leps}{\Lmu}
\end{equation}
is composed of
\begin{equation}
\left.\begin{aligned}
{\Heps}&=\nt^2+\frac{\epsz\mut}{\dt}\left(\pzz+\kt^2\right)\\
{\Hmu}&=\nt^2+\frac{\epst\muz}{\dt}\left(\pzz+\kt^2\right)\\
{\Leps}&=-\frac{\epsz\xit}{\dt}\left(\pzz+\kt^2\right)\\
{\Lmu}&=-\frac{\muz\xit}{\dt}\left(\pzz+\kt^2\right)\\
\end{aligned}\right\}\,,
\end{equation}
where
$\nabla^2\equiv\nt^2+\partial^2/\partial z^2$, $\dt=\epst\mut-\xit^2$, and $\kt^2=\omega^2\dt$. This operator
was factorized by Weiglhofer \cite{W1994} as
\begin{equation}
\label{factor}
\calH=\calH_+\calH_-\,,
\end{equation}
where
\begin{equation}
\calH_\pm=\nt^2+A_\pm\left(\pzz+\kt^2\right)
\end{equation}
and
\begin{equation}
A_\pm=\frac{1}{2\dt}\left\{\epst\muz+\epsz\mut\mp\left[(\epst\muz-\epsz\mut)^2+4\epsz\muz\xit^2\right]^{1/2}\right\}\,.
\end{equation}
The square root in $A_\pm$ is taken on the principal branch. Its real part is nonnegative and, when that real part vanishes, its imaginary part is nonnegative. Reversing its sign merely interchanges the labels $+$ and $-$. We assume $A_+\ne A_-$ throughout.

The factorization \eqref{factor} of $\calH$ is crucial for the initiation of the scalar factorization method \cite{LW1994,WL1994a,WL1994b}.
 Given Eq.~\eqref{eq9},
 $W(\xx,\xp)$ satisfies  both of the following relations:
\begin{equation}
\label{Wprop}
\left.\begin{array}{l}
\displaystyle{
\nt^2W(\xx,\xp)=-\frac{1}{A_+-A_-}\les{A_+G_+(\xx,\xp)-A_-G_-(\xx,\xp)}\ris }\\[5pt]
\displaystyle{
\ler \pzz+\kt^2\rir W(\xx,\xp)=\frac{1}{A_+-A_-}\les{G_+(\xx,\xp)-G_-(\xx,\xp)}\ris }
\end{array}\right\}\,.
\end{equation}
Here, the scalar Green functions  
\begin{equation}
G_\pm(\xx,\xp)=\frac{\exp(i\kt D_\pm)}{4\pi D_\pm}
\end{equation}
with
\begin{equation}
\left.\begin{array}{l}
D_\pm=+\sqrt{A_\pm\rho^2+(z-z')^2}\\[5pt]
\rho=+\sqrt{(\xx-\xp)^2-(z-z')^2}
\end{array}\right\}\,
\end{equation}
satisfy the differential equations \cite{L1992}
\begin{equation}
\calH_\pm G_\pm(\xx,\xp)=-\delta(\xx-\xp)\,.
\end{equation}
Here $\rho$ is the nonnegative real square root. For the outgoing Green functions with the $\exp(-i\omega t)$ convention, the branches of $\kt=\omega\sqrt{\dt}$ and $D_\pm$ are fixed by continuation from $\omega>0$, $\dt>0$, and $A_\pm>0$, where $\kt>0$ and $D_\pm>0$ away from the source. For complex constitutive parameters, the continuation is taken with the outgoing limiting-absorption prescription, so that the choices of $\kt$ and $D_\pm$ must  not be made independently.

The adjoint operator for the TUBM with equal magnetoelectric scalars can be written as
\begin{equation}
\label{def-La123}
\La(\nabla)=\=L_1\nt^2+\=L_2\left(\pzz+\kt^2\right)+\=L_3,
\end{equation}
where
\begin{subequations}
\begin{equation}
\label{def-L1}
\=L_1=-\frac{1}{\kt^2}\left[\mut\left(\nabla\nabla+\omega^2\epst\epsz\mut
\=\eps^{-1} \right)+\omega^2\xit^2(\muz\It-\mut\uv z\uv z)\right]\,,
\end{equation}
\begin{equation}
\label{def-L2}
\begin{aligned}
\=L_2&=-\frac{1}{\kt^2}\bigg\{ \muz\left(\nabla\nabla+\omega^2\epst\epsz\mut 
\=\eps^{-1}\right)-\omega^2\muz\xit^2\uv z\uv z \\
&
\qquad+i\omega\muz\xit\les\cz\uv z+\uv z\cz\ris\bigg\} \,,
\end{aligned}
\end{equation}
and
\begin{equation}
\label{def-L3}
\begin{aligned}
\=L_3&=\frac{\ce}{\dt}\cz\cz\\
&\qquad-\frac{i\muz\xit}{\omega\dt}\frac{\partial}{\partial z}\bigl[\nt\cz+\cz\nt\bigr]\,,
\end{aligned}
\end{equation}
\end{subequations}
with $\nt =\nabla - ( \partial / \partial z) \uv z $.
 
On making use of Eqs.~\eqref{def-La},  \eqref{Wprop}, \eqref{def-La123}, and \eqref{prop1}, we obtain
\begin{equation}
\label{def-Gee}
\G_{ee}(\xx,\xp)=\G_+(\xx,\xp)+\G_-(\xx,\xp)+\F(\xx,\xp),
\end{equation}
where 
\begin{equation}
\label{prop1}
\G_\pm(\xx,\xp)=\mp\les{A_\pm\=L_1-\=L_2}\ris \frac{G_\pm(\xx,\xp)}{\dA}\,
 \end{equation}
and
\begin{equation}
\F(\xx,\xp)=\=L_3W(\xx,\xp)\,.
\end{equation}
Following Ref.~\citenum{W1994}, we split up $\F(\xx,\xp)$  as
\begin{equation}
\label{def-F123}
\F(\xx,\xp)=\F_1(\xx,\xp)+\F_2(\xx,\xp)+\F_3(\xx,\xp)\,,
\end{equation}
where
\begin{subequations}
\begin{equation}
\label{def-F1}
\F_1(\xx,\xp)=\frac{\ce}{\dt}\cz\cz W(\xx,\xp) \,,
\end{equation}
\begin{equation}
\label{def-F2}
\F_2(\xx,\xp)=-\frac{i\muz\xit}{\omega\dt}\frac{\partial}{\partial z}\bigl[\nt\cz+\cz\nt\bigr]W(\xx,\xp)\,,
\end{equation}
and 
\begin{equation}
\label{def-F3}
\F_3(\xx,\xp)=\=L_3W(\xx,\xp)-\les \F_1(\xx,\xp)+\F_2(\xx,\xp)\ris\,.
\end{equation}
\end{subequations}

Weiglhofer \cite{W1994} treated $\F_3(\xx,\xp)$ as the solution of a fourth-order equation of the type 
\begin{equation}
\label{eq26}
\calH\F_3=\=Q\delta(\xx-\xp)\,, 
\end{equation}
where he had $\=Q\ne\=0$ erroneously.
He could not solve   Eq.~\eqref{eq26} in closed form, which  led him to the conclusion that
$\G_{ee}(\xx,\xp)$ cannot be determined in closed form.
In contrast, Eqs.~\eqref{def-L1}--\eqref{def-L3}
show  that $\=L_3$ consists of exactly the two terms that produce $\F_1(\xx,\xp)$ and $\F_2(\xx,\xp)$. In other words, $\=Q=\=0$ so that
\begin{equation}
\label{F3=zero}
\F_3(\xx,\xp)=\=0\, .
\end{equation}

Since every term on the right sides of Eqs.~\eqref{def-F1} and \eqref{def-F2} differentiates $W(\xx,\xp)$ with respect to the transverse coordinates but  $W(\xx,\xp)$
depends on $\rho$ and $z-z'$ only, the first of Eqs.~\eqref{Wprop} can be integrated once with respect to $\rho$. With the integration constant fixed by regularity on the axis $\rho=0$, we get
\begin{equation}
\label{Wprop1}
\nt W(\xx,\xp)= \uv\rho\frac{i\tau}{4\pi\kt\dA\rho}\,,
\end{equation}
where
\begin{equation}
\tau=\exp(i\kt D_+)-\exp(i\kt D_-)\,
\end{equation}
and $(\uv\rho,\uv\phi,\uv z)$ are the unit vectors of the cylindrical coordinate system.
In addition,
\begin{equation}
\label{Wprop2}
\cz W(\xx,\xp)=\les\nt W(\xx,\xp)\ris\times\uv z\,.
\end{equation}
 Hence $\F_1(\xx,\xp)$ and $\F_2(\xx,\xp)$ are available in closed form, and no part of $\G_{ee}(\xx,\xp)$ remains undetermined. Of course,
 all results here hold in the sense of distributions. Furthermore, terms in $\rho^{-2}$ are understood through their limits on the axis $\rho=0$ \cite{W1994}.

\section{Dyadic Green functions in closed form}
\subsection{Electric field due to electric current}
Equations~\eqref{def-Gee}, \eqref{def-F123}, and \eqref{F3=zero} deliver
\begin{equation}
\label{Gee-final}
\G_{ee}(\xx,\xp)=\G_+(\xx,\xp)+\G_-(\xx,\xp)+\F_1(\xx,\xp)+\F_2(\xx,\xp),
\end{equation}
where
\begin{subequations}
\begin{equation}
\begin{aligned}
\label{G-final}
\G_\pm(\xx,\xp)=&\pm\frac{1}{\kt^2\dA}\Big\{ (\mut A_\pm-\muz)\left(\nabla\nabla+\omega^2\epst\epsz\mut\=\eps^{-1}\right)\\
&+\omega^2\xit^2\les A_\pm\muz\It+(\muz-A_\pm\mut)\uv z\uv z \ris\\
&-i\omega\muz\xit\les \cz\uv z+\uv z\cz\ris \Big\} G_\pm(\xx,\xp)\,,
\end{aligned}
\end{equation}
\begin{equation}
\label{F1-final}
\F_1(\xx,\xp)= \frac{\ce}{\dt\dA}\,\M(\xx,\xp)\,,
\end{equation}
and
\begin{equation}
\label{F2-final}
\F_2(\xx,\xp)=-\frac{i\muz\xit}{\omega\dt\dA}\frac{\partial}{\partial z}
\Bigl[\uv z\times\M(\xx,\xp)-\M(\xx,\xp)\times\uv z\Bigr]\,,
\end{equation}
\end{subequations}
with
\begin{equation}
\label{def-M}
\M(\xx,\xp)=
(\It-2\uv\phi\uv\phi)\frac{i\tau}{4\pi\kt\rho^2}-[A_+G_+(\xx,\xp)-A_-G_-(\xx,\xp)]\uv\phi\uv\phi\,.
\end{equation}
Equations~\eqref{Gee-final}--\eqref{def-M} were the ultimate goal of Weiglhofer \cite{W1994}.

\subsection{Magnetic field due to electric current}
With $\bm K=\bm 0$, the first of Eqs.~\eqref{Maxwellcurl} gives 
\begin{equation}
\label{def-Gme}
\G_{me}=\=\mu^{-1}\. (\curlI-i\omega\=\zeta)\. \G_{ee}\,.
\end{equation}
Accordingly,
\begin{equation}
\G_{me}(\xx,\xp)=\G_{me+}(\xx,\xp)+\G_{me-}(\xx,\xp)+\F_{me}(\xx,\xp)\,,
\end{equation}
where
\begin{subequations}
\begin{equation}
\begin{aligned}
\G_{me\pm}(\xx,\xp)&=\pm\frac{1}{\kt^2\dA}\,\=\mu^{-1}\. \Bigg\{ -i\omega\xit(\mut A_\pm-\muz)\nt\nabla\\
&+\omega^2\les \epsz\mut(\mut A_\pm-\muz)+\muz\xit^2A_\pm\ris (\curlI-i\omega\xit\It)\\
&+\omega^2\les\mut(\epst-\epsz)(\mut A_\pm-\muz)-(\mut+\muz)\xit^2A_\pm\ris \cz\uv z\\
&-i\omega\muz\xit\les \frac{\partial}{\partial z}\nabla\uv z-\uv z\uv z\nabla^2+\cz\cz\ris\Bigg\} G_\pm(\xx,\xp)
\end{aligned}
\end{equation}
and
\begin{equation}
\begin{aligned}
\F_{me}(\xx,\xp)&=\=\mu^{-1}\. \Bigg\{ \frac{\ce}{\dt}\Bigl[\Bigl(\frac{\partial}{\partial z}\nabla-\uv z\nabla^2\Bigr)\cz-i\omega\xit\cz\cz\Bigr]\\
& \quad +\frac{i\muz\xit}{\omega\dt}\frac{\partial}{\partial z}\Bigl[\frac{\partial}{\partial z}\cz\cz-\Bigl(\frac{\partial}{\partial z}\nabla-\uv z\nabla^2\Bigr)\nt\\
&\hspace{4mm}+i\omega\xit\bigl(\nt\cz+\cz\nt\bigr)\Bigr]\Bigg\}W(\xx,\xp)\,.
\end{aligned}
\end{equation}
\end{subequations}
In the expression for $\F_{me}(\xx,\xp)$, $W(\xx,\xp)$ appears only through $\nt W(\xx,\xp)$ and $\cz W(\xx,\xp)$, which are available in closed form from 
Eqs.~\eqref{Wprop1} and \eqref{Wprop2}.

\subsection{Magnetic field due to magnetic current}
The duality transformation $\bm E\to\bm H$, $\bm H\to-\bm E$, $\bm J\to\bm K$, $\bm K\to-\bm J$ \cite{L1992} preserves 
Eqs.~\eqref{Maxwellcurl} for  the TUBM with equal magnetoscalars provided that
 $\eps_{t,z}\leftrightarrow\mu_{t,z}$ and $\xit\to-\xit$. This transformation leaves $\dt$, $\kt$, $A_\pm$, $G_\pm$, $\tau$ and $W$ unchanged.
Hence, $\G_{mm}$ is obtained from $\G_{ee}$ as
\begin{equation}
\G_{mm}(\xx,\xp)=\G_{mm+}(\xx,\xp)+\G_{mm-}(\xx,\xp)+\E_{1}(\xx,\xp)+\E_{2}(\xx,\xp),
\end{equation}
where
\begin{subequations}
\begin{equation}
\begin{aligned}
\G_{mm\pm}(\xx,\xp)=&\pm\frac{1}{\kt^2\dA}\Big\{(\epst A_\pm-\epsz)\left(\nabla\nabla+\omega^2\mut\muz\epst\=\mu^{-1}\right)\\
&+\omega^2\xit^2\les (A_\pm\epsz\It+(\epsz-A_\pm\epst)\uv z\uv z\ris \\
&+i\omega\epsz\xit\les \cz\uv z+\uv z\cz\ris \Big\} G_\pm(\xx,\xp)\,,
\end{aligned}
\end{equation}
\begin{equation}
\label{E1-final}
\E_{1}(\xx,\xp)=\frac{\cm}{\dt\dA}\,\M(\xx,\xp)\,,
\end{equation}
and
\begin{equation}
\label{E2-final}
\E_{2}(\xx,\xp)=\frac{i\epsz\xit}{\omega\dt\dA}\frac{\partial}{\partial z}
\Bigl[\uv z\times\M(\xx,\xp)-\M(\xx,\xp)\times\uv z\Bigr]\,.
\end{equation}
\end{subequations}

\subsection{Electric field due to magnetic current}
With $\bm J=\bm 0$, the second of Eqs.~\eqref{Maxwellcurl} gives $\G_{em}=-\=\eps^{-1}\. (\curlI+i\omega\=\xi)\. \G_{mm}$, which is also the dual of $\G_{me}$. Hence,
\begin{equation}
\G_{em}(\xx,\xp)=\G_{em+}(\xx,\xp)+\G_{em-}(\xx,\xp)+\F_{em}(\xx,\xp),
\end{equation}
where
\begin{subequations}
\begin{equation}
\begin{aligned}
\G_{em\pm}(\xx,\xp)=&\mp\frac{1}{\kt^2\dA}\, \=\eps^{-1}\. \Bigg\{ i\omega\xit(\epst A_\pm-\epsz)\nt\nabla\\
&+\omega^2\les \epst\muz(\epst A_\pm-\epsz)+\epsz\xit^2A_\pm\ris (\curlI+i\omega\xit\It)\\
&+\omega^2\les\epst(\mut-\muz)(\epst A_\pm-\epsz)-(\epst+\epsz)\xit^2A_\pm\ris \cz\uv z\\
&+i\omega\epsz\xit\les \frac{\partial}{\partial z}\nabla\uv z-\uv z\uv z\nabla^2+\cz\cz\ris\Bigg\} G_\pm(\xx,\xp)
\end{aligned}
\end{equation}
and
\begin{equation}
\begin{aligned}
\F_{em}(\xx,\xp)=&-\=\eps^{-1}\. \Bigg\{\frac{\cm}{\dt}\Bigl[\Bigl(\frac{\partial}{\partial z}\nabla-\uv z\nabla^2\Bigr)\cz+i\omega\xit\cz\cz\Bigr]\\
&-\frac{i\epsz\xit}{\omega\dt}\frac{\partial}{\partial z}\Bigl[\frac{\partial}{\partial z}\cz\cz-\Bigl(\frac{\partial}{\partial z}\nabla-\uv z\nabla^2\Bigr)\nt\\
&-i\omega\xit\bigl(\nt\cz+\cz\nt\bigr)\Bigr]\Bigg\}W(\xx,\xp)\,.
\end{aligned}
\end{equation}
\end{subequations}

\section{Concluding remarks}
All four dyadic Green functions of the TUBM with equal magnetoelectric scalars are expressed through $G_\pm$, $\tau$, and $\rho$ as well
as their derivatives, provided $\dt\ne0$, $A_\pm\ne0$ and $A_+\ne A_-$. Written as in Eqs.~\eqref{F2-final}
 and \eqref{E2-final}, $\F_2(\xx,\xp)$ and $\E_{2}(\xx,\xp)$ remain valid when the coefficients of $\F_1(\xx,\xp)$ and $\E_{1}(\xx,\xp)$ vanish. For $\xit=0$ the results reduce to those for a uniaxial dielectric-magnetic medium \cite{W1990}.  
 
 With this success of the scalar factorization method, which had unfortunately eluded Weiglhofer \cite{W1994}
 because of some derivational errors, we expect this method to produce more novel results soon.

\vspace{10mm}
 
\section*{Acknowledgments}
Both authors were supported  by EPSRC (grant number APP68512).   

\section*{Declaration of competing interest}
The authors declare that they have no known competing financial interests or personal relationships that could have
appeared to influence the work reported in this paper.

\section*{Declaration of AI use}
This paper emerged while experimenting with ChatGPT-6.1~Sol~Light to implement the scalar factorization method.
All equations were typeset by that chatbot. The authors wrote the textual material and verified the equations.

\section*{Data availability}
All data  analyzed in this paper are included in this paper.

\end{document}